\documentclass{article}
\usepackage{array} 
\usepackage{amsthm}
\usepackage{amsmath}
\usepackage{amsfonts}
\usepackage{amssymb}
\usepackage{graphicx} 
\usepackage{mathtools}
\usepackage{stmaryrd}
\usepackage{xcolor}
\usepackage{dsfont}
\usepackage{hyphenat}
\usepackage{hyperref}
\usepackage{forest}
\usepackage{tikz}
\usetikzlibrary{arrows,patterns}
\usetikzlibrary {arrows.meta,bending,positioning}
\usepackage{accents}
\usepackage{titlesec}
\usepackage{caption} 
\usepackage{nicematrix} 

\usepackage[backend=bibtex,style=alphabetic]{biblatex} 
\newcommand\avrg[1]{\langle#1\rangle}
\newcommand\bra[1][]{\langle#1|}
\newcommand\configurations{\mathcal{G}}
\newcommand\erf{\textrm{erf}}
\newcommand\evolution{\mathcal{E}}
\newcommand\evol{E}
\newcommand\Id{\mathbb{I}} 
\newcommand\intrange[1][]{\llbracket #1\rrbracket} 
\newcommand\ket[1][]{|#1\rangle} 
\newcommand\N{\mathbb{N}}
\newcommand\op[1][]{\hat{#1}} 

\newcommand\recur{\mathcal{R}}
\newcommand\state{\mathcal{S}} 
\newcommand\zcfg[1]{|#1\rangle}

\newtheorem{corollary}{Corollary}
\newtheorem{lemma}{Lemma}
\newtheorem{prop}{Proposition}

\newtheorem{conjecture}{Conjecture}

\title{Space-time correlations in the $1$D Directed Stochastic Sandpile model}
\author{Valentin Lallemant$^1$\\
$^1$\small{\textit{Université Grenoble Alpes, CNRS, LPMMC, 38000 Grenoble, France}}}
\date{December 2025}

\begin{document}

\maketitle
\begin{abstract}
    Sandpile models are known to resist exact results.
    In this direction, space-time correlations between avalanches have proven to be especially difficult to access.
    One of the main obstacle to do so comes from taking memory effects in a systematic way along the computation.
    In this paper, we partially fill this gap and derive recursive relations for the particle filling and avalanche $2$-points correlation function in the $1$D Directed Stochastic Sandpile.
    These expressions allow to characterize the sign of the correlations and estimates are provided in the particle filling case.
    In fact, density correlations are shown to be positively correlated.
    This behavior is directly related to persistence of the local particle filling.
    On the other hand, we show that avalanches are anticorrelated in the model.
    This is interpreted by the fact that avalanches disrupt the system and the damage can only be fully compensated after injecting a sufficiently high number of particles.
    These results indicate an underlying trade off, between static and dynamic observable, for the system to sit in its stationary state.
    It appears that this balance is controlled by the conservation of the particle number along the avalanches.
\end{abstract}
\newpage

\section{Introduction}

Self-organised criticality (SOC) is a versatile concept which has now matured on phenomenological and exact grounds. 
SOC was introduced as an attempt to capture generic features of an out of equilibrium system to generate spontaneous power laws \cite{bak_self-organized_1987}.
In this respect, it has been shown, among the last $4$ decades of research, that the so called class of sandpile models incorporates the phenomenology of the originally conceived SOC (see for example the thesis \cite{fajardo_universality_2008}).

Sandpile models can be defined by a set of general rules. 
Consider a space discrete system in which particles evolve. 
In this system, particle movements are triggered by a local observable such as the local number of particles (on site for example).
In fact, particles can accumulate at each site as long as they stay below a model specific local filling threshold. 
When the threshold is exceeded, the local configuration gets unstable and particles are redistributed in the neighborhood of the denser region. 
After a while, the system relaxes, redistributing the particles until, eventually, the dynamical process ends.
At this point, the system has reached a stable configuration. 
To develop a stationary state, a driving is imposed upon the system, which basically consist in the addition of particles at a given frequency, and dissipative boundaries are specified so that particles can leave.
In fact, the injection of particles in the system must be sufficiently slow so that the relaxation process is intermittent. 
Each relaxation event is then called an avalanche.
One way to enforce this separation of time scale between the rate of injection and the dissipative events consist in injecting one particle only after the previous avalanche has faded. 
Equivalently, one can set a rate of injection so small (scaling with the system size) that it is compatible with the previous statement.
On another note, local conservation of the particle number along the redistribution has been shown to be quite central for the emergence of a SOC state \cite{fajardo_universality_2008}.

In these sandpile models, a lot of efforts have been put toward characterizing their stationary state. 
In fact, even accessing the stationary state expression can be, and is usually, a great challenge. 
Rigorous results in this direction might culminate with the Abelian Sandpile Model (ASM) in $2$ space dimensions, yet only covers static observable (e.g. density, density-density space correlations) or immediate response of the system to a perturbation (e.g. avalanche waves probability distribution). 
A certain number of these result can be found in the review \cite{jarai_sandpile_2018}. 
Despite this achievements, correlations characterising the static properties (density) or response (avalanche size) of the system in time, meaning between successive avalanches, have been largely out of reach, leaving a hole in the comprehension of the sandpile models at the level of rigorous results.

In this paper, we hope to fill partially this gap and bring answers to open problems on sandpile models (see e.g. \cite{jonsson_area_1998, dhar_theoretical_2006}). 
We are precisely concerned with space-time correlations in one of the simplest sandpile model with a stochastic dynamics in $1D$, which we call the Directed Stochastic Sandpile (DSS).
Notice that other types of correlations, which should connect to the result of the present paper, have been studied around variations of the DSS \cite{welinder_multiscaling_2007,garcia-millan_correlations_2018}.
In fact, the DSS is quite general and was introduced in different contexts, mostly gravitating around the concept of SOC.
The model was probably first properly defined in \cite{pruessner_exact_2004} as the Totally Asymmetric Oslo Model (TAOM). 
As the name suggests, it was conceived as a simplified version of the Oslo model \cite{christensen_tracer_1996}, another $1$D sandpile model in which particles can flow in both directions.
The simplification in this context finds its origin in forcing the flow of particles to occur only in one of the direction. 
In a similar way, the DSS was introduced in the literature of the Activated Random Walk (ARW) model \cite{dickman_activated_2010}, and was again seen as a simplified model, because of its directedness.
The DSS model was also generalised and studied using algebraic methods in \cite{alcaraz_directed_2008}.
Finally, DSS was incompletely described in \cite{jonsson_area_1998}, where a $1$D Random Walk (RW) with an absorbing wall problem was studied. 
Indeed, in their article, the authors believed that this RW problem would model the propagation of an avalanche front, which revealed latter to be true \cite{pruessner_exact_2004}.
In fact, this mapping between this RW problem and the DSS is a central tool in its analysis as will be shown in the next sections.

After introducing formally the model, we gather a number of important results in the driven-dissipative stationary state setup. 
We continue and find a recursive relation for the density $2$-points correlation function, and unveil some of its symmetries.
We perform a similar, yet more involved, analysis on the avalanche $2$-points correlation function, and also find a recursive relation plus some additional properties of this object. 
We end the discussion with a summary of the results and some open questions left along the analysis.
\section{Definitions}
We start with a formal definition of the Directed Stochastic Sandpile (DSS) model.
The model describes the evolution of particles on a linear chain of $L$ sites.
We represent the system at a given time by a state of generalised configurations. 
We denote by $\configurations$ the set of generalised configurations and $\state:=\textrm{span}(\configurations)$ the vector space defined as 
\begin{equation}
    \state = \big\{\sum_{i=1}^{k}\lambda_ic_i~|~\forall i~\lambda_{i}\geq 0 ,~\sum_{i=1}^{k}\lambda_i=1,~c_i\in\configurations,~ k<\infty \big\}
\end{equation}
A configuration $c\in\configurations$ is made of a pair of discrete fields $c\sim(w,z)$, where $z(x)\in\{0,1\}$ and $w(x)\in\N$ are respectively describing the number of stable and waiting particles on site $x\in\intrange[1,L]$. 
For convenience, we will usually adopt the following configuration representation
\begin{equation}
    (w,z)=\prod_{x=1}^{L}a_x^{w(x)}\zcfg{z(1),...,z(L)}
\end{equation}
where $a_x$ represents exactly one waiting particle on site $x$.

Now, as already mentioned, the system's particles have the ability to evolve.
In fact, stable particles do not evolve whereas waiting particles can perform different instructions: either they jumps to the next site to their right, either they become stable, either they convert stable particles into waiting particles.
To be precise, the local evolution rules are completely defined by the two relations
\begin{subequations}\label{seteq:rule DSS}
\begin{align}
    a_x\zcfg{1}&\to a_x a_{x-1}\zcfg{0}\\
    a_x\zcfg{0}&\to p\zcfg{1}+qa_{x+1}\zcfg{0}
\end{align}
\end{subequations}
where $p$ and $q$ give respectively the probabilities for a waiting particle to stabilize, if $z(x)=0$, or to jump to the right site.
We call from now on the jump instruction a toppling.
Obviously, $p+q=1$ and $p,q>0$.
Remark that there are no interactions \emph{between waiting particles}.
Moreover, we say that a configuration $c\sim(w,z)$ is stable if the condition $w(x)=0$ is satisfied for all $x$. 

For completeness, we introduce a set of evolution operators $\evolution$ where the action of each $\evol\in\evolution$ applied on a $c\in\configurations$ corresponds to evolve \emph{exactly one waiting particle} of each configuration of the initial state.
Besides, for each pair $\evol,\evol'\in\evolution$ where $\evol\neq\evol'$, there must exist $c\in\configurations$ for which $\evol(c)\neq \evol(c')$, meaning they choose different sites with a waiting particle to be evolved.
The set of equations \eqref{seteq:rule DSS} is then equivalent to 
\begin{subequations}\label{eq: rules DSS}
\begin{align}
    \evol(a_x\zcfg{1})&= a_x a_{x-1}\zcfg{0}\\
    \evol(a_x\zcfg{0})&= p\zcfg{1}+qa_{x+1}\zcfg{0}
\end{align}
\end{subequations}
In this paper, we are only concerned with \emph{the driven-dissipative} setup, where waiting particles are added at $x=1$ once the system is stable, and particles leave the system at $L+1$, meaning $a_{L+1}\phi=\phi$ for any $\phi\in\state$.
We call $\recur$ the set of recurrent configurations, configurations in which the system has a non zero probability to appear in the stationary state, and so the stationary state $\psi$ is an element of $\textrm{span}(\recur)$.
\section{State of the art}\label{section: estimates}
We condense a number of exact results known on the model \cite{jonsson_area_1998, pruessner_exact_2004, stapleton_one-dimensional_2006, alcaraz_directed_2008} adapted with our choice of notations.
A general strategy, which has proven efficient and is extensively used in the present paper, relies on a mapping between the avalanches and the area enclosed by a $1D$ random walker evolving in a space where an absorbing wall is settled.
This approach was notably formalised and developed in \cite{pruessner_exact_2004}.

\subsection{Preliminaries}
First of all, the dynamics described by \eqref{eq: rules DSS} is abelian.
\begin{prop}[Stabilisation is Abelian, \cite{pruessner_exact_2004}]\label{prop: DSS is Abelian}
    For any $\evol,\evol'\in \evolution$ with $\evol\neq\evol'$, and any initial state $\phi\in\state$, we have the equality
    \begin{equation}
        \evol^{\infty}(\phi)=\evol'^{\infty}(\phi)
    \end{equation}
    where $\evol^k:=\evol\circ\evol\circ... \evol$ is the iterated composition of the operator.
\end{prop}
This proposition implies that the order in which one evolves the waiting units has no consequences on the final stable state.
We are therefore free to choose the most convenient one depending on the context.

\begin{prop}[Stationary state, \cite{pruessner_exact_2004}]\label{prop: NESS DSS}
    The stationary state $\psi$ of the driven-dissipative setup is a product state given by 
    \begin{equation}\label{eq: NESS DSS}
        \psi = \bigotimes_{x=1}^L \big(p\zcfg{1_x} + q\zcfg{0_x}\big)
    \end{equation}
    where $\zcfg{...1_x...}$ notation means "site $x$ has $1$ particle" (e.g. $\zcfg{1,0,1,0,1}=\zcfg{1_1,0_2,1_3,0_4,1_5}$) and $\bigotimes$ is the iterated tensor product.
    The recurrent set $\recur$ is therefore equal to $\{0,1\}^L$.
    Moreover, writing $W:=\evol^{\infty}\circ a_1$, we have for any initial state $r\in\recur$ and for $\tau=\min(L,L+1-\sum_{x=1}^{L}z^r(x))$ the equality
    \begin{equation}\label{eq: reset to NESS}
        W^\tau(r)=\psi
    \end{equation}
\end{prop}
For the rest of the paper, $\psi$ will always denote the stationary state \eqref{eq: NESS DSS}. 

\subsection{Avalanches and a random walker with one absorbing wall}
Following the ideas of \cite{pruessner_exact_2004}, we now introduce the mapping of the avalanches in the stationary state with a RW problem.
Propositions \ref{prop: DSS is Abelian} and \ref{prop: NESS DSS} justify to conceive the avalanches as the propagation of a front.
The front is a site where all waiting particles of a configuration are accumulated.
Because the dynamics is directed, once the front passes site $x$, the local density at $x$ is fixed and cannot be disturbed anymore by the ongoing avalanche.
For example, consider an initial state $\zcfg{1,0,1}$ with $L=3$. 
Upon the injection of a particle at $x=1$, we have 
\begin{align*}
    a_1\zcfg{1,0,1} &\to pa_2\zcfg{1,0,1}+qa_2^2\zcfg{0,0,1}\\
    &\to p(p\zcfg{1,1,1}+qa_3\zcfg{1,0,1})+q(pa_3\zcfg{0,1,1}+qa_3^2\zcfg{0,0,1})\\
    &\to  p(p\zcfg{1,1,1}+q(pa_4\zcfg{1,0,1}+qa_4^2\zcfg{1,0,0}))\\
    &+q(p(pa_4\zcfg{0,1,1}+qa_4^2\zcfg{0,1,0})+q(pa_4^2\zcfg{0,0,1}+qa_4^3\zcfg{0,0,0}))
\end{align*}
At each evolution step described above, the front propagates from its site to the next one on the right for all the configurations containing at least one waiting particle.
The last state obtained is stable as $a_4:=1$ (boundary condition of the driven-dissipative setup).
The sequence of states that we compute when propagating the front is therefore just a subsequence of the states generated after iteration of one well-chosen operator $\evol\in\evolution$ on the initial configuration.

\begin{figure}[ht]
    \centering
    \includegraphics[width=0.7\linewidth]{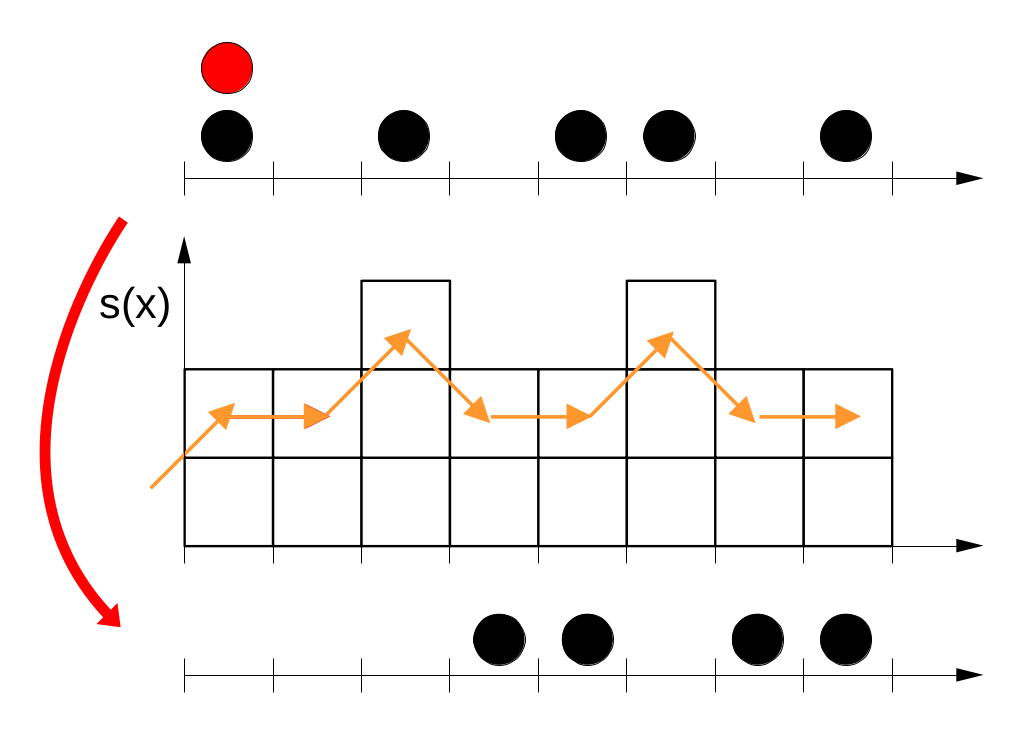}
    \caption{Transition between $a_1\zcfg{1,0,1,0,1,1,0,1}$ and $\zcfg{0,0,0,1,1,0,1,1}$.
    The avalanche $s(x)$ connecting the two configurations is explicited in the center of the figure.}
    \label{fig: ex transition}
\end{figure}

Denote by $s\in\N$ the size of an avalanche involving $s$ topplings.
In particular, we can decompose the avalanche as a sum of correlated random variables $s:=\sum_{x=1}^Ls(x)$ where $s(x)$ corresponds to the number of toppling at site $x$ along the avalanche or, equivalently, the number of waiting particles transferred from site $x$ to site $x+1$.
In the stationary state, the question of the avalanche probability distribution $P(s)$ can be reformulated \cite{pruessner_exact_2004} in terms of a random walker whose position corresponds to the height of the front of waiting particles. 
We show in Figure \ref{fig: ex transition} an example of a transition between two configurations, with the corresponding avalanche specified.

\begin{figure}[ht]
    \centering
    \includegraphics[width=0.6\linewidth]{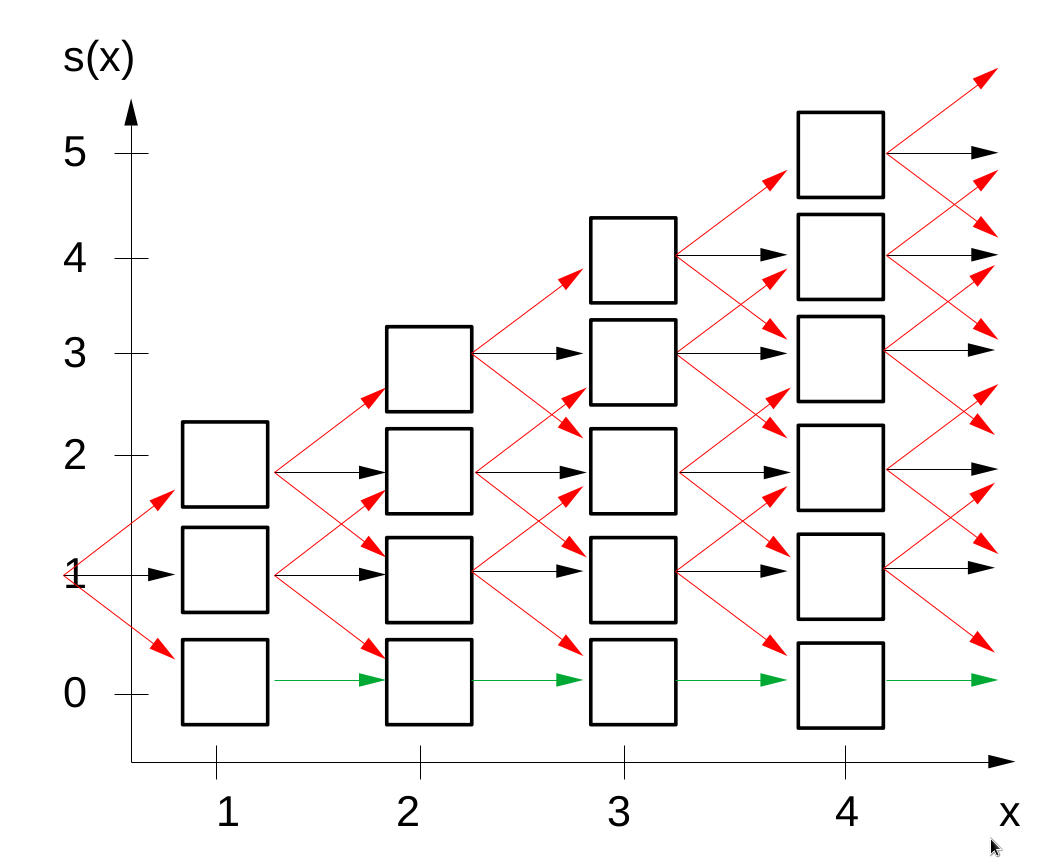}
    \caption{Random walker picture of the avalanches in the stationary state of the DSS. 
    Red and black arrows have respective probabilities $\alpha=pq$ and $\beta=1-2\alpha$ of occuring.
    The green arrows are deterministic processes.
    Each square represents a position of the random walker.}
    \label{fig: RW picture}
\end{figure}

Now, recalling that we are in the stationary state $\psi$, we can conceive the initial condition $a_1^\tau\psi$ equivalently as a random walker starting at a height $\tau$ and moving toward the right. 
Denote by $P(y|\tau,x)$ the probability of the $1$D random walker to go from $(\tau,0)\longrightarrow (y,x)$, where the first and second indices are respectively the RW position (height of the particle front, vertical axis), and the time of its evolution (site of the chain, horizontal axis), knowing that an absorbing boundary is set at the height $0$.
In the stationary case, $P(y|\tau,x)$ satisfies the set of equation
\begin{subequations}\label{eq: RW recursion}
    \begin{align}
        P(y|\tau,x+1) &= \alpha_{+}P(y-1|\tau,x)+\alpha_{-}P(y+1|\tau,x)+\beta P(y|\tau,x)~,\forall y\geq 2\\
        P(1|\tau,x+1) &= \alpha_{-}P(2|\tau,x)+\beta P(1|\tau,x)\\
        P(0|\tau,x+1) &= \alpha_{-}P(1|\tau,x)+P(0|\tau,x)\label{eq: abs wall recurence}
    \end{align}
\end{subequations}
where $\alpha_{-},\beta$ and $\alpha_{+}$ correspond respectively to the probabilities for the movements $-1,0$ and $+1$ of the random walker. 
These probabilities are related to the microscopic parameters of the original DSS model by 
\begin{equation*}
    \alpha_{+}=\alpha_{-}=pq,~\beta = 1-\alpha_{+}-\alpha_{-}=1-2pq
\end{equation*}
The transitions depicted in \eqref{eq: RW recursion} are easily pictured with the RW problem. 
A geometrical representation of these recursions is provided in Figure \ref{fig: RW picture}.
Notice also that, given that we start in the stationary state, the random walker performs a symmetric walk. 

\subsection{Solutions and estimates around the random walk formulation}

From this RW construction, one would like to come up with an explicit solution for $P(y|\tau,x)$.
This problem can and has already been solved along several lines.
Here, using the \emph{method of images}, one can express $P(y|\tau,x)$, for all $y\geq 1$ ($y=0$ case is treated next), as the difference of the probability distributions of one free (no absorbing wall) random walker starting at the position $\tau$ and one at $-\tau$, evaluated at the same "time" and height $x$ and $y$. 
In equation, this reads
\begin{align}
    P^{\textrm{free}}(y|\tau,x) &=\sum_{n=0}^{\lfloor\frac{L-|y-\tau|}{2}\rfloor}\alpha_{+}^{|y-\tau|+n}\alpha_{-}^{n}\beta^{L-|y-\tau|-2n}\binom{L}{|y-\tau|+n,n}
\end{align}
where $P^{\textrm{free}}(y|\tau,x)$ is the probability of one free random walker to perform the transition $(\tau,0)\longrightarrow (y,x)$. 
The solution for $y\geq 1$ is then given by 
\begin{equation}\label{eq: images method}
    P(y|\tau,x)=P^{\textrm{free}}(y|\tau,x)-P^{\textrm{free}}(y|-\tau,x)
\end{equation}
In the next sections, we will be interested in the probability $P(0|\tau,x)$ of the RW, or the avalanche, to stop propagating before or at site $x$.
This quantity can be expressed in many ways
\begin{equation}\label{eq: all expressions absorbing probabilty RW} 
    P(0|\tau,x) = 1-\sum_{y=1}^{L+1}P(y|\tau,x)=1-\sum_{y=1}^{2\tau}P^{\textrm{free}}(y|\tau,x)=\alpha_{-}\sum_{x'=0}^{x-1}P(1|\tau,x')
\end{equation}
The first equality uses the normalisation condition of the pdf, the second equality develops and re sums the solution in terms of the free RW problem \eqref{eq: images method}. 
For the third equality, we use the boundary condition \eqref{eq: abs wall recurence}.
As a side remark, from a combinatorial point of view, the question of determining $P(1|\tau,x')$ is directly related to weighted variations around the Motzkin numbers \url{https://oeis.org/A001006}.
This can be seen from Figure \ref{fig: RW picture}.

Now, applying different techniques on this RW representation, it was shown that
\begin{prop}[Scaling form of $P(s)$, \cite{jonsson_area_1998, pruessner_exact_2004}]\label{prop: P(s) scaling}
    The avalanche probability density function $P(s)$ has the scaling form
    \begin{equation}
        P(s)\sim s^{-\tau}f\bigg(\frac{s}{s^*}\bigg)
    \end{equation}
    where $\tau =\frac{4}{3}$ and $s^*\sim L^D$ is the cutoff of the power law regime with $D=\frac{3}{2}$. 
    More, the scaling function $f$ behaves as 
    \begin{equation}
        f\bigg(\frac{s}{s^*}\bigg)\left\{\begin{array}{lc}
            \sim 1 & \text{for}~s<s^* \\
            \leq C_1\exp(-C_2\frac{s}{s^*}) & \text{for}~s>s^*
        \end{array}
        \right.
    \end{equation}
    with $C_1$ and $C_2$ some positive constants.
\end{prop}

Using the RW picture, we can also provide an estimate for the scaling form of $P(0|\tau,x)$
\begin{prop}[estimate on $P(0|\tau,x)$]\label{prop: bound P(0|tau,x)}
    For sufficiently large $x$, we have the following estimate on the probability $P(0|\tau,x)$ for a random walker starting at the position $\tau>0$ to be absorbed before or at the $x^{th}$ time steps by an horizontal absorbing wall settled at the height $y=0$
    \begin{equation}
        P(0|\tau,x) \left\{\begin{array}{ll}
            \sim 1-\erf\bigg(\frac{\tau}{\sqrt{2Cx}}\bigg)& ,\forall x \geq \tau \\
            =0 & ,\forall x \leq \tau-1 
        \end{array}
        \right.
    \end{equation}
    Here, $\sigma^2 \sim Cx$, for some $C>0$ and $\mu = \tau$ are respectively the dominant scaling of the variance and the average position of a free random walker (no absorbing wall) starting at the position $\tau$, after $x$ time steps.
\end{prop}

\begin{proof}
    In the context of the RW literature, given that $1<<x$, i.e. in the long time regime, the problem converges to a Brownian motion. 
    Given the pdf $P^{\textrm{free}}(y|\tau, x)$, we have a typical variance $\sigma^2 \sim Cx$, for some $C>0$, and average position $\mu = \tau$ of the RW.
    We have the estimate
    \begin{equation*}
        \sum_{y=a}^b P^{\textrm{free}}(y|\tau, x) \sim \int_{y=a}^{b} (2\pi Cx)^{-\frac{1}{2}}\exp\big(-\frac{(y-\tau)^2}{2Cx}\big)dy
    \end{equation*}
    Using \eqref{eq: all expressions absorbing probabilty RW} we get
    \begin{align*}
        P(0|\tau,x) &=1-\sum_{y=1}^{2\tau}P^{\textrm{free}}(y|\tau,x)= 1-\sum_{y=-\tau+1}^{\tau}P^{\textrm{free}}(y|0,x)\\
        &\sim 1-\int_{y=-\tau}^{\tau} (2\pi Cx)^{-\frac{1}{2}}\exp\bigg(-\frac{y^2}{2Cx}\bigg)dy\\
        &\sim 1-\erf\bigg(\frac{\tau}{\sqrt{2Cx}}\bigg)
    \end{align*}
    where $\erf(z):=\frac{2}{\sqrt{\pi}}\int_{0}^z\exp(-t^2)dt$ is the error function.
\end{proof}
Last but not least, we mention and adapt a result proven for the Oslo model \cite{dhar_steady_2004}, another $1$D stochastic sandpile.
The statement says that for scalar observable which depend only on the stable distribution, time correlations between time $t$ and $t+\tau$, i.e. stable distributions \emph{after} the $t^{th}$ and $(t+\tau)^{th}$ avalanches, are exactly $0$ for all time greater than a finite time $\tau^*(L)$.
Since the DSS is a close relative of the Oslo model, the same property can be verified.
\begin{prop}[Finite time correlations, \cite{dhar_steady_2004}]\label{prop: finite time correlations}
    Consider an operator $\hat{A}$ probing a scalar observable of a state.
    Then, starting from the stationary state $\psi$ at time $t$, the time correlation reads 
    \begin{equation}\label{eq: finite time correlations}
        \avrg{A(t)A(t+\tau)}-\avrg{A(t)}\avrg{A(t+\tau)} = 0
    \end{equation}
    for all $\tau\geq L$.
    In the case where $\hat{A}$ probes a transport property between two successive states $\phi$ and $W(\phi)$, we have the same result as \eqref{eq: finite time correlations}, starting from $\psi$, but valid this time for $\tau\geq L+1$.
\end{prop}
\begin{proof}
    The case for the scalar observable can be solved with the exact same arguments as in \cite{dhar_steady_2004}, so we decided to focus on the case where $\hat{A}$ probes a transport property (such as the size of the avalanches). 
    $\hat{A}$ can always be represented by a square matrix so that we have rewrite
    \begin{align*}
        \avrg{A(t)A(t+\tau)}-\avrg{A(t)}\avrg{A(t+\tau)} = \bra[\Id]\op[A]W^{\tau-1}\op[A]\ket[\psi]-\bra[\Id]\op[A]\ket[\psi]\bra[\Id]\op[A]W^{\tau-1}\ket[\psi]
    \end{align*}
    where $\bra[\Id]:= \bigotimes_{x=1}^L \big(\bra[1_x] + \bra[0_x]\big)=\sum_{r\in\recur}\bra[r]$ and $\bra[r]\ket[r']=\delta_{r,r'}$. 
    The notation $W^{\tau}$ corresponds to the matrix $W$ at the power $\tau$.
    The time in the model is given by the number of avalanches, and there are $\tau-1$ intermediate ones that separate the initial avalanche triggered by the operator $\hat{A}$ on the right of $\bra[\Id]\op[A]W^{\tau-1}\op[A]\ket[\psi]$ and the $\tau^{th}$ one with the $\hat{A}$ on the left.
    Notice that 
    \begin{align*}
        W^{\tau-1}\op[A]\ket[\psi] = W^{\tau-1}\sum_{r,r'\in\recur}A(r\to r')P(r)\zcfg{r'}= \sum_{r,r'\in\recur}A(r\to r')P(r) W^{\tau-1}\zcfg{r'}
    \end{align*}
    When $\tau-1\geq L$, we know from \eqref{eq: reset to NESS} that 
    \begin{align*}
       \sum_{r,r'\in\recur}A(r\to r')P(r) W^{\tau-1}\zcfg{r'} =\big(\sum_{r,r'\in\recur}A(r\to r')P(r)\big) \ket[\psi] =C\ket[\psi]
    \end{align*}
    Remark that the above equation is false for any $\tau-1< L$ taking for instance the term $W^{\tau-1}\zcfg{0^L}$.
    In the end, when $\tau-1\geq L$, i.e. $\tau \geq L+1$ we get
    \begin{align*}
        \bra[\Id]\op[A]W^{\tau-1}\op[A]\ket[\psi] = C\bra[\Id]\op[A]\ket[\psi] = C^2
    \end{align*}
    Injecting this results in the original expression, one gets $\avrg{A(t)A(t+\tau)}-\avrg{A(t)}\avrg{A(t+\tau)} = C^2-C^2 = 0$.
\end{proof}
\section{The particle filling is persistent}
We investigate the $2$-points correlation of the particle filling.
To shorten the notations, we denote equivalently by $z(x,t)$ and $z_x^t$ the realisation of the RV which counts the number of particles at site $x$ \emph{after} $t$ avalanches starting from the stationary state $\psi$.
\begin{prop}\label{prop:persistence z(x,t) z(x,t+tau)}
In the stationary state, the $2$-points correlation function of the local particle filling $z(x,t)$ satisfies the following conditions for all $x\in\intrange[1,L]$
\begin{equation}\label{eq: z(x,t) is persistent}
    \avrg{z(x,t)z(x,t+\tau)} -\avrg{z(x,t)}\avrg{z(x,t+\tau)} \left\{\begin{array}{cl}
        >0 &\text{if}~ 1\leq \tau \leq x-1\\
        =0 & \text{if}~ \tau\geq x
    \end{array}
    \right.
\end{equation}
More, in the regime $1\leq \tau\leq x-1$, the following relation is verified
\begin{equation}\label{eq: scaling correlations z(x,t)}
    \avrg{z(x,t)z(x,t+\tau)}-\avrg{z(x,t)}\avrg{z(x,t+\tau)}
    =P(0|\tau,x-1)pq
\end{equation}
where $P(0|\tau,x)$ is defined in Proposition \ref{prop: bound P(0|tau,x)}.
\end{prop}
\begin{proof}
By definition, we have
\begin{align*}
     \avrg{z(x,t)z(x,t+\tau)} &= \sum_{z_x^t,~z_x^{t+\tau}}z_x^t z_x^{t+\tau}P(z_x^t\cap z_x^{t+\tau})\\
     &=P(z_x^t=1\cap z_x^{t+\tau}=1)\\
     &= P(z_x^{t+\tau}=1|z_x^t=1)P(z_x^t=1)
\end{align*}
The event $z_x^t=1$ happens with probability $p$ at time $t$ since we are in the stationary state. 
It only remains to deal with $P(z_x^{t+\tau}=1|z_x^t=1)$, which is the probability of having $z_x^{t+\tau}=1$ knowing that $\tau$ avalanches have been performed since $z_x^t=1$.

We know that the stationary state has no spatial correlations as its a product state (see Proposition \ref{prop: NESS DSS}).
At time $t$ we are then only concerned with the state 
\begin{equation*}
    \varphi = \bigg(\bigotimes_{y=1}^{x-1}(p\zcfg{1_y}+q\zcfg{0_y})\bigg)\zcfg{1_x}\bigg(\bigotimes_{y=x+1}^{L}(p\zcfg{1_y}+q\zcfg{0_y})\bigg)
\end{equation*}
from which the $\tau$ avalanches will be initiated.
As we have already discussed, the abelianess of the model allows to consider the $\tau$ avalanches as one  super avalanche triggered after the addition of $a_1^{\tau}$ to $\varphi$.
This problem is mapped to a RW problem starting at the position $\tau$ and with an absorbing wall at the position $0$. 
From the definition of $\varphi$ and previous results (cf Figure \ref{fig: RW picture}), the random walk is symmetric until it reaches site $x$ where it is biased, as we are not locally in the stationary state.

There are now only two possible events: either the super avalanche stops before reaching $x$, and $z_{x}^{t+\tau}=1$ deterministically since we had $z_{x}^{t}=1$; either at least one waiting particle is transferred on site $x$, and there is a probability $p$ for a particle to stay at $x$ after the front has passed (independently of the front size). 
They define a Bernoulli RV $\epsilon$ with a probability of success $\theta := 1-P(0|\tau,x-1)$ (success event $\epsilon=1$) for the super avalanche front to pass by $x$, and of failure $1-\theta$  ($\epsilon=0$) for the front to stop before site $x$.
We can then rewrite 
\begin{align*}
    \avrg{z(x,t)z(x,t+\tau)} &= P(z_x^{t+\tau}=1|z_x^t=1)P(z_x^t=1)\\
    &= p(P(z_x^{t+\tau}=1|z_x^t=1\cap\big(\epsilon =0\cup \epsilon =1\big)))\\
    &= p(P(z_x^{t+\tau}=1|z_x^t=1\cap\epsilon =0) + P(z_x^{t+\tau}=1|z_x^t=1\cap\epsilon =1))\\
    &=p((1-\theta)+\theta p)
\end{align*}
This gives the relation \eqref{eq: scaling correlations z(x,t)}
\begin{equation*}
    \avrg{z(x,t)z(x,t+\tau)}-\avrg{z(x,t)}{z(x,t+\tau)} =p((1-\theta)+\theta p)-p^2=P(0|\tau,x-1)pq
\end{equation*}
This quantity is positive for any $p\in]0,1[$ and any $P(0|\tau,x-1)>0$.
The latter condition, and so the inequality \eqref{eq: z(x,t) is persistent}, are verified as long as $\tau\leq x-1$. 
The case where $\tau\geq x$ is treated in Proposition \ref{prop: finite time correlations}.
\end{proof}

Proposition \ref{prop:persistence z(x,t) z(x,t+tau)} states that there is \emph{persistence} of the particle filling in the system. 
It originates from a trade-off. 
Avalanches do not spread in the system with the same scaling as the system size, but we know from the stationary condition that we must dissipate at least one particle at the boundary. 
To compensate this behavior, there exists large dissipative avalanches.
Now, for all the configurations where the super avalanche have stopped before the sink, all the sites at the right of the last site with a toppling are necessarily unchanged and is in fact the only contribution giving rise to a non zero particle filling correlations.
This result connects in an exact way the avalanches and bulk properties of the stationary state.
Remark also that \eqref{eq: scaling correlations z(x,t)} is bounded from above by $pq$. 
Strikingly, there is also a non trivial symmetry of \eqref{eq: scaling correlations z(x,t)} which evaluates to the same value when swapping the microscopic probabilities $p\leftrightarrow q$.

We can now extend Proposition \ref{prop:persistence z(x,t) z(x,t+tau)} to different sites $x$ and $x'$
\begin{prop}\label{prop: zero correlations z(x,t) z(x',t')}
For all $x \neq x'$, we have
\begin{equation}\label{eq:corr z(x,t,y,t+tau)}
   \avrg{z(x,t) z(x',t+\tau)} -\avrg{z(x,t)}\avrg{z(x',t+\tau)} = 0
\end{equation}
\end{prop}

\begin{proof}
Suppose $x>x'$.
Since the dynamics is directed, i.e. exploiting the causal order of the avalanches, the events $z_x^t=1$ and $z_{x'}^{t+\tau}$ are independent, and so $P(z_x^t=1\cap z_{x'}^{t+\tau}=1)= P(z_x^t=1)P(z_{x'}^{t+\tau}=1)$ and
\begin{align*}
    \avrg{ z(x,t) z(x',t+\tau) } -\avrg{z(x,t)}\avrg{z(x',t+\tau)}= p^2-p^2 =0
\end{align*}
Suppose now $x<x'$.
Using the same decomposition as in the previous proof, we get
\begin{align*}
    \avrg{ z(x,t) z(x',t+\tau) } &= P(z_x^t=1)P(z_{x'}^{t+\tau}=1|z_x^t=1) \\
    &= p\big[P(z_{x'}^{t+\tau}=1\cap \epsilon = 0|z_x^t=1)+P(z_{x'}^{t+\tau}=1\cap \epsilon = 1|z_x^t=1)\big] \\
    &= p\big[P(z_{x'}^{t+\tau}=1|z_x^t=1\cap \epsilon = 0)P( \epsilon = 0)+\\
    &~P(z_{x'}^{t+\tau}=1|z_x^t=1\cap \epsilon = 1)P( \epsilon = 1)\big]\\
    &= p\big[p(1-\theta)+p\theta\big]= p^2
\end{align*}
where $\epsilon$ is the Bernoulli RV which says if the super avalanche front ($a_1^\tau$) have reached the site $x'$ (success $\epsilon=1$ with probability $\theta$) or not, \emph{conditioned on the fact that} $z_x^t=1$. 
We used also $P(z_{x'}^{t+\tau}=1|z_x^t=1\cap \epsilon = 0)=P(z_{x'}^t)=p$ as we start from the stationary state.
Subtracting $\avrg{z(x,t)}\avrg{z(x',t+\tau)}=p^2$ to $\avrg{ z(x,t) z(x',t+\tau) }$ gives the desired result.
\end{proof}

Proposition \ref{prop: zero correlations z(x,t) z(x',t')} establishes that in the stationary state, the filling of particles between sites are uncorrelated in time.
This extends one result of \cite{pruessner_exact_2004} where the product form of the stationary state was already ensuring no space correlation at equal time.  
We conclude from the previous propositions that persistence only happens at equal sites. 
This extends naturally at the mesoscopic and macroscopic level
\begin{prop}\label{prop: persistance global density}
The total filling $z(t):=\sum_{x=1}^{L}z(x,t)$ is persistent, in the sens that
\begin{equation}\label{eq: persistence z(t)}
    \avrg{z(t)z(t+\tau)}-\avrg{z(t)}\avrg{z(t+\tau)}\left\{
    \begin{array}{lc}
    > 0 & \text{for}~1\leq \tau \leq L-1\\
    = 0 & \text{for}~\tau\geq L
    \end{array}
    \right.
\end{equation}
with the relation
\begin{equation}\label{eq: estimate correlations z(t)}
    \avrg{z(t)z(t+\tau)}-\avrg{z(t)}\avrg{z(t+\tau)} = pq\sum_{x=\tau+1}^{L}P(0|\tau,x-1)
\end{equation}
valid for all $1\leq \tau \leq L-1$
\end{prop}
\begin{proof}
By definition, we have
\begin{align*}
    \avrg{ z(t) z(t+\tau) } &= \avrg{ \sum_x z(x,t)\sum_{x'}z(x',t+\tau)}\\
    &= \sum_{r,r'\in\recur^2} P(r)P(a_1^{\tau}r\to r') \sum_{x,x'} z^r(x)z^{r'}(x')\\
    &= \sum_{x,x'}\sum_{r,r'\in\recur^2}P(r)P(a_1^{\tau}r\to r') z^r(x)z^{r'}(x') \\
    &= \sum_{x,x'}\avrg{z(x,t)z(x',t+\tau)}
\end{align*}
The double sum contains $L^2$ terms, and we know that $\avrg{z(t)}=\avrg{z(t+\tau)}=pL$ as we start and stay in the stationary states.
We have therefore 
\begin{align*}
     \avrg{z(t)z(t+\tau)} -\avrg{z(t)}\avrg{z(t+\tau)} &=\bigg(\sum_x \sum_{x'}\avrg{z(x,t)z(x',t+\tau)}\bigg) -p^2L^2\\
     &=\sum_x \sum_{x'}\big(\avrg{z(x,t)z(x',t+\tau)} -p^2\big)\\
     &=\sum_{x=\tau+1}^{L}\big(\avrg{z(x,t)z(x,t+\tau)} -\avrg{z(x,t)}\avrg{z(x,t+\tau)}\big) >0
\end{align*}
where we used Propositions \ref{prop:persistence z(x,t) z(x,t+tau)} and \ref{prop: zero correlations z(x,t) z(x',t')}.
To get \eqref{eq: estimate correlations z(t)} it suffices to inject \eqref{eq: scaling correlations z(x,t)} in the last sum $\sum_{x=\tau+1}^{L}...$.
\end{proof}
\section{Avalanches are anticorrelated}

In this section, given an avalanche at time $t$ involving $s(t)$ topplings, we denote equivalently by $s(x,t)=s_x^t$ the number of local topplings at site $x$ of this avalanche. 
By construction, $s(t):=\sum_{x=1}^Ls(x,t)=\sum_{x=1}^Ls_x^t=s^t$.
As for the previous section, these notations will become particularly handful as the expressions get heavier.
We should stress that the coming results are a bit more involved than the one on the density.
It arises from the necessity to implement recursive relations in space, i.e. expressing correlations at a site compared to the same quantity at an equal time but on the nearest neighbour site on the left.
We start with an underlying symmetry of the avalanche correlations.
\begin{prop}[Min property of $\avrg{s(x,t)s(x',t+\tau)}$]\label{prop: min s(x,t) s(x',t+tau)}
Suppose the system to be initially in the stationary state.
Then the following equality holds for all $x,x'$
\begin{equation}\label{eq: min corr s(x,t) s(x',t+1)}
    \avrg{s(x,t)s(x',t+\tau)} = \avrg{s(l,t)s(l,t+\tau)}
\end{equation}
where $l=\min(x,x')$.
\end{prop}
\begin{proof}
For convenience we start with $\tau = 1$ and will deal with the general case $\tau \geq 1$ along the argumentation.
Suppose first the case where $x'>x$.
We have
\begin{align*}
    \avrg{s(x,t) s(x',t+1)} &= \sum_{s_x^t=1}^{x+1}\sum_{s_{x'}^{t+1}=1}^{x'+1}s_x^t s_{x'}^{t+1}P(s_x^t\cap s_{x'}^{t+1}) \\
    P(s_x^t\cap s_{x'}^{t+1})&=P(s_x^t)P(s_{x'}^{t+1}|s_x^t) 
\end{align*}
For convenience, for any $x$ and $t$, we will use the shorthand $\sum_{s_x^t}s_x^t...:=\sum_{s_x^t=1}^{x+1}s_x^t...$ for the summation as the term proportional to $s_x^t = 0$ does not contribute.
In the case where the distribution on site $x'$ is, conditionally on $s_x^t$, the stationary distribution, i.e. $p\zcfg{1_{x'}}+q\zcfg{0_{x'}}$ after the $t^{th}$ avalanche, we have the relation (see the left sketch of Figure \ref{fig:CV s(x') s(x'-1)})
\begin{multline}\label{eq:min prop s(x,t) s(x',t+1)}
P(s_{x'}^{t+1}|s_x^t)=\alpha_{+}P(s_{x'-1}^{t+1}=s_{x'}^{t+1}-1|s_x^t)+\alpha_{-}P(s_{x'-1}^{t+1}=s_{x'}^{t+1}+1|s_x^t)\\
+\beta P(s_{x'-1}^{t+1}=s_{x'}^{t+1}|s_x^t)
\end{multline}
Now, after the avalanche $s^t$, we show that all sites $x''\geq x+1$ are locally in the stationary distribution. 
This is obvious since
\begin{equation}\label{eq: NESS on x'>x knowing s(x,t)}
    a_{x+1}^{s_x^t}\bigotimes_{i=x+1}^{L}(p\zcfg{1_i}+q\zcfg{0_i})\to \bigotimes_{i=x+1}^{L}(p\zcfg{1_i}+q\zcfg{0_i})
\end{equation}
for any $s_x^t\geq 0$. 
This means that Equation \eqref{eq:min prop s(x,t) s(x',t+1)} holds for all $x'\geq x+1$.
Remark that this property holds also for any $\tau\geq 1$, and so does \eqref{eq:min prop s(x,t) s(x',t+1)}.
Indeed, in between the first (time $t$) and the $\tau^{th}$ avalanche (at time $t+\tau$), the $t'\in\intrange[1,\tau-1]$ avalanches in between will contribute exactly in the same way as in Equation \eqref{eq: NESS on x'>x knowing s(x,t)} but replacing $s_x^t$ by $s_{x}^{t'}$.
The argument that follows can then be easily adapted to the general case $\tau\geq 1$.

Now, it suffices to perform a change variable from $s_{x'}^{t+1}\to s_{x'-1}^{t+1}$ (see Figure \ref{fig:CV s(x') s(x'-1)} for a graphical representation) to conclude
\begin{align*}
     \avrg{s(x,t) s(x',t+1)} &= \sum_{s_x^t}s_x^t P(s_x^t)\sum_{s_{x'}^{t+1}}s_{x'}^{t+1}P(s_{x'}^{t+1}|s_x^t)\\
     &= \sum_{s_x^t}s_x^t P(s_x^t)\sum_{s_{x'}^{t+1}}s_{x'}^{t+1}\big[\alpha_{+}P(s_{x'-1}^{t+1}=s_{x'}^{t+1}-1|s_x^t)\\
     &+\alpha_{-}P(s_{x'-1}^{t+1}=s_{x'}^{t+1}+1|s_x^t)+\beta P(s_{x'-1}^{t+1}=s_{x'}^{t+1}|s_x^t)\big]\\
     &=\sum_{s_x^t} s_x^t P(s_x^t)\sum_{s_{x'-1}^{t+1}=1}^{x'}\big(\alpha_{+}(s_{x'-1}^{t+1}+1)\\
     &+\alpha_{-}(s_{x'-1}^{t+1}-1)+\beta s_{x'-1}^{t+1}\big)P(s_{x'-1}^{t+1}|s_x^t)\\
     &=\sum_{s_x^t}s_x^t P(s_x^t)\sum_{s_{x'-1}^{t+1}}s_{x'-1}^{t+1}P(s_{x'-1}^{t+1}|s_x^t)\\
     &=:\avrg{s(x,t) s(x'-1,t+1)}
\end{align*}
since $\alpha_{+}=\alpha_{-}$ and $\alpha_{+}+\alpha_{-}+\beta=1$ .
We can iterate the construction from site $x'-1$ until we reach site $x$ which proves Equation \eqref{eq: min corr s(x,t) s(x',t+1)} when $x'>x$. 
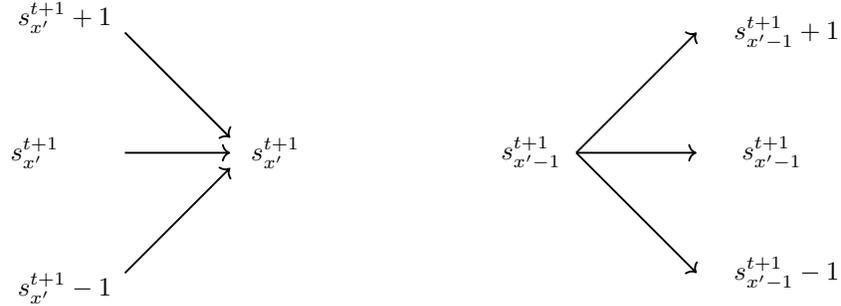
\begin{figure}[h!]
    \centering
    \begin{tikzpicture}
    \begin{scope}[scale=0.8]
        \draw[line width=0.25mm, ->] (1,0.25) -- (2.75,0.25);
        \draw[line width=0.25mm, ->] (1,2.25) -- (2.75,0.5);
        \draw[line width=0.25mm, ->] (1,-1.75) -- (2.75,0);
        \node at (3.5,0.25) {$s_{x'}^{t+1}$};
        \node at (0,2.5) {$s_{x'}^{t+1}+1$};
        \node at (-0.5,0.25) {$s_{x'}^{t+1}$};
        \node at (0,-2) {$s_{x'}^{t+1}-1$};
    \end{scope}
    \begin{scope}[xshift = 6cm, scale=0.8]
        \draw[line width=0.25mm, ->] (1,0.25) -- (3,0.25);
        \draw[line width=0.25mm, ->] (1,0.25) -- (3,2.25);
        \draw[line width=0.25mm, ->] (1,0.25) -- (3,-1.75);
        \node at (0.25,0.25) {$s_{x'-1}^{t+1}$};
        \node at (4.25,0.25) {$s_{x'-1}^{t+1}$};
        \node at (4.5,2.25) {$s_{x'-1}^{t+1}+1$};
        \node at (4.5,-1.75) {$s_{x'-1}^{t+1}-1$};
    \end{scope}
    
    \end{tikzpicture}
    \caption{(Left) Change of Variable (CV) fixing $s_{x'}^{t+1}$ when the local state at $x'$ is $p\zcfg{1_{x'}}+q\zcfg{0_{x'}}$ just after the avalanche at time $t$.
    (Right) Equivalent CV, but fixing instead $s_{x'-1}^{t+1}$.
    Ascending, horizontal and descending arrows happen respectively with the probabilities $\alpha_{+},~\beta$ and $\alpha_{-}$}
    \label{fig:CV s(x') s(x'-1)}
\end{figure}

Suppose now $x'<x$. 
If \eqref{eq: min corr s(x,t) s(x',t+1)} holds, then, in particular, $\avrg{s(x,t)s(x,t+1)}$ must be equal to $\avrg{s(x+1,t)s(x,t+1)}$.
For any fixed $s_{x+1}^t,$ and $s_x^{t+1}$, we have
\begin{align*}
    P(s_{x+1}^t\cap s_x^{t+1}) &=P(\big(\bigcup_{\forall s_x^t}s_x^t\big)\cap s_{x+1}^t\cap s_x^{t+1})\\
    &=\sum_{\forall s_x^t}P(s_x^t\cap s_{x+1}^t\cap s_x^{t+1})\\
    &=\sum_{\forall s_x^t}P(s_{x+1}^t\cap s_x^{t+1}|s_x^t)P(s_x^t)\\
    &=\sum_{\forall s_x^t}P(s_x^{t+1}|s_x^t\cap s_{x+1}^t)P(s_{x+1}^t|s_x^t)P(s_x^t)\\
    &=P(s_x^{t+1}|s_x^t=s_{x+1}^t-1)\alpha_{+}P(s_x^t=s_{x+1}^t-1)\\
    &+P(s_x^{t+1}|s_x^t=s_{x+1}^t+1)\alpha_{-}P(s_x^t=s_{x+1}^t+1)\\
    &+P(s_x^{t+1}|s_x^t=s_{x+1}^t)\beta P(s_x^t=s_{x+1}^t)
\end{align*}
where we used the conditional independence $P(s_x^{t+1}|s_x^t\cap s_{x+1}^t)=P(s_x^{t+1}|s_x^t)$ since $s_x^{t+1}$ depends only on the profile $z(x')$ with $x'\leq x$, so only on $s^x_t$, as the dynamic is directed. 
Again, this property holds for arbitrary $\tau\geq 1$ so that the proof can be generalised quite directly.
We can now perform the CV $s_{x+1}^t\to s_x^t$
\begin{align*}
    \avrg{s(x+1,t)s(x,t+1)} &= \sum_{s_{x+1}^t,~s_x^{t+1}}s_{x+1}^t s_x^{t+1} P(s_{x+1}^t\cap s_x^{t+1}) \\
    &=\sum_{s_{x+1}^t,~s_x^{t+1}}s_{x+1}^t s_x^{t+1}\big[P(s_x^{t+1}|s_x^t=s_{x+1}^t-1)P(s_x^t=s_{x+1}^t-1)\alpha_{+}\\
    &+P(s_x^{t+1}|s_x^t=s_{x+1}^t+1)P(s_x^t=s_{x+1}^t+1)\alpha_{-}\\
    &+P(s_x^{t+1}|s_x^t=s_{x+1}^t)P(s_x^t=s_{x+1}^t)\beta\big] \\
    &=\sum_{s_x^t,~s_x^{t+1}} s_x^{t+1}\big[(s_x^t+1)P(s_x^{t+1}|s_x^t)P(s_x^t)\alpha_{+}\\
    &+(s_x^t-1)P(s_x^{t+1}|s_x^t)P(s_x^t)\alpha_{-}+s_x^tP(s_x^{t+1}|s_x^t)P(s_x^t)\beta\big]\\
    &= \sum_{s_x^t,~s_x^{t+1}} s_x^{t+1}s_x^tP(s_x^{t+1}|s_x^t)P(s_x^t) \\
    &=\avrg{s(x,t)s(x,t+1)}
\end{align*}
By induction on $\avrg{s(x',t)s(x',t+1)}$, we can reach $\avrg{s(x,t)s(x',t+1)}$ for any $x'<x$, and \eqref{eq: min corr s(x,t) s(x',t+1)} is consequently true.
\end{proof}

\begin{lemma}\label{lemma: <s(x,t)s(x,t+1)> decreases with x}
    $\avrg{s(x,t)s(x,t+\tau)}$ is a decreasing function with increasing $x$.
    Precisely, we have the recursion 
    \begin{multline}\label{eq: precise recursion <s(x,t)s(x,t+tau)>}
        \avrg{s(x,t)s(x,t+\tau)} = \avrg{s(x-1,t)s(x-1,t+\tau)}\\
        -\alpha \sum_{s_{x-1}^{t},~s_{x-1}^{t+\tau}=1}^{x}(1-\theta(s_{x-1}^t,\tau-1))P(s_{x-1}^{t+\tau}|s_{x-1}^{t}\cap \epsilon=0)P(s_{x-1}^{t})
    \end{multline}
    where $\epsilon$ is a Bernoulli RV with probability of success $\theta(s_{x-1}^t,\tau-1)$ which corresponds to the event where the $\tau-1$ avalanches between the initial avalanche and the $\tau^{th}$ one have send at least one particle on site $x$. 
    This RV depends obviously on the first avalanche through $s_{x-1}^t$, and the number of intermediate avalanches $\tau-1$.
\end{lemma}
\begin{proof}
    For simplicity, we start setting $\tau = 1$ and will argue for the general $\tau \geq 1$ case at the end of the demonstration. 
    We use the same approach as in Proposition \ref{prop: min s(x,t) s(x',t+tau)}, meaning we want to perform the change of variable from $s_x^t\to s_{x-1}^t$
    \begin{align*}
        \avrg{s(x,t)s(x,t+1)}&=\sum_{s_x^{t},s_x^{t+1}}s_x^{t}s_x^{t+1}P(s_x^{t}\cap s_x^{t+1})\\
        &=\sum_{s_x^{t+1}}s_x^{t+1}\sum_{s_x^{t},~s_{x-1}^{t}}s_x^{t} P(s_{x-1}^{t}\cap s_{x}^{t} \cap s_x^{t+1})\\
        &=\sum_{s_x^{t+1}}s_x^{t+1}\sum_{s_x^{t},~s_{x-1}^{t}}s_x^{t} P(s_x^{t+1}|s_{x-1}^{t}\cap s_{x}^{t})P(s_{x}^{t}|s_{x-1}^{t})P(s_{x-1}^{t})
    \end{align*}
    and we have 
    \begin{align*}
        \sum_{s_x^{t},~s_{x-1}^{t}}s_x^{t}& P(s_x^{t+1}|s_{x-1}^{t}\cap s_{x}^{t})P(s_{x}^{t}|s_{x-1}^{t})P(s_{x-1}^{t})\\
        &=\sum_{s_{x-1}^{t}}(s_{x-1}^{t}+1)P(s_x^{t+1}|s_{x-1}^{t}\cap s_{x}^{t}=s_{x-1}^{t}+1)\alpha_{+}P(s_{x-1}^{t})\\
        &+(s_{x-1}^{t}-1)P(s_x^{t+1}|s_{x-1}^{t}\cap s_{x}^{t}=s_{x-1}^{t}-1)\alpha_{-}P(s_{x-1}^{t})\\
        &+s_{x-1}^{t}P(s_x^{t+1}|s_{x-1}^{t}\cap s_{x}^{t}=s_{x-1}^{t})\beta P(s_{x-1}^{t})\\
        &=\sum_{s_{x-1}^{t}}s_{x-1}^{t}P(s_x^{t+1}\cap s_{x-1}^{t})\\
        &+\alpha \sum_{s_{x-1}^{t}}\big[P(s_x^{t+1}|s_{x-1}^{t}\cap s_{x}^{t}=s_{x-1}^{t}+1)-P(s_x^{t+1}|s_{x-1}^{t}\cap s_{x}^{t}=s_{x-1}^{t}-1)\big]P(s_{x-1}^{t})
    \end{align*}
    where we used 
    \begin{align*}
        P(s_x^{t+1}\cap s_{x-1}^{t}) &= P(s_x^{t+1}\cap s_{x-1}^{t}\cap\bigcup_{\delta^t\in\{-1,0,1\}}(s_{x}^t=s_{x-1}^{t}+\delta^t))\\
        &=P(s_x^{t+1}|s_{x-1}^{t}\cap s_{x}^{t}=s_{x-1}^{t}+1)\alpha_{+}P(s_{x-1}^{t})\\
        &+P(s_x^{t+1}|s_{x-1}^{t}\cap s_{x}^{t}=s_{x-1}^{t}-1)\alpha_{-}P(s_{x-1}^{t})\\
        &+P(s_x^{t+1}|s_{x-1}^{t}\cap s_{x}^{t}=s_{x-1}^{t})\beta P(s_{x-1}^{t})
    \end{align*}
    which is true as long as $s_{x-1}^t>0$.
    This is indeed the case as the $s_{x-1}^t=0$ does not contribute by construction to the sum.
    Therefore 
    \begin{multline}\label{eq: recursion <s(x,t)s(x,t+1)>}
        \avrg{s(x,t)s(x,t+1)} = \avrg{s(x-1,t)s(x-1,t+1)}\\
        +\alpha \sum_{s_{x-1}^{t},~s_x^{t+1}}s_x^{t+1} \big[P(s_x^{t+1}|s_{x-1}^{t}\cap s_{x}^{t}=s_{x-1}^{t}+1)\\
        -P(s_x^{t+1}|s_{x-1}^{t}\cap s_{x}^{t}=s_{x-1}^{t}-1)\big]P(s_{x-1}^{t})
    \end{multline}
    where we used the Proposition \ref{prop: min s(x,t) s(x',t+tau)} for the first term in the r.h.s..
    For the statement made in proposition to hold, we must have 
    \begin{equation}\label{eq: condition of decreasing}
       \sum_{s_{x-1}^{t},~s_x^{t+1}}s_x^{t+1} \big[P(s_x^{t+1}|s_{x-1}^{t}\cap s_{x}^{t}=s_{x-1}^{t}+1)
        -P(s_x^{t+1}|s_{x-1}^{t}\cap s_{x}^{t}=s_{x-1}^{t}-1)\big]P(s_{x-1}^{t})<0
    \end{equation}
    To prove this, we make the CV $s_x^{t+1}\to s_{x-1}^{t+1}$.
    We have 
    \begin{align*}
        P(s_x^{t+1}|s_{x-1}^{t}&\cap s_{x}^{t}=s_{x-1}^{t}+1)= \sum_{s_{x-1}^{t+1}}P(s_{x-1}^{t+1}\cap s_{x}^{t+1}|s_{x-1}^{t}\cap s_{x}^{t}=s_{x-1}^{t}+1)\\
        &=\sum_{\delta^{t+1}\in\{-1,0,1\}}P(s_{x}^{t+1}|s_{x-1}^{t}\cap s_{x}^{t}=s_{x-1}^{t}+1\cap s_{x-1}^{t+1}=s_x^{t+1}-\delta^{t+1})\times\\
        &P(s_{x-1}^{t+1}=s_x^{t+1}-\delta^{t+1}|s_{x-1}^{t}\cap s_{x}^{t}=s_{x-1}^{t}+1)
    \end{align*}
    and 
    \begin{subequations}\label{eq: set conditional prob tau=1}
    \begin{align}
        P(s_{x}^{t+1}|s_{x-1}^{t}\cap s_{x}^{t}=s_{x-1}^{t}+1\cap s_{x-1}^{t+1}=s_x^{t+1}-\delta^{t+1})=\left\{\begin{array}{cc}
            p &  \text{if}~\delta^{t+1}= -1 \\
            q &  \text{if}~\delta^{t+1}= 0 \\
            0 &  \text{if}~\delta^{t+1}= 1
        \end{array}\right.\\
        P(s_{x}^{t+1}|s_{x-1}^{t}\cap s_{x}^{t}=s_{x-1}^{t}-1\cap s_{x-1}^{t+1}=s_x^{t+1}-\delta^{t+1})=\left\{\begin{array}{cc}
            0 &  \text{if}~\delta^{t+1}= -1 \\
            p &  \text{if}~\delta^{t+1}= 0 \\
            q &  \text{if}~\delta^{t+1}= 1
        \end{array}\right.
    \end{align}
    \end{subequations}
    Indeed, remark that after the avalanche with $s_{x-1}^{t}$, we have necessarily $z(x,t^+)=1,0$ respectively for $\delta^t:=s_x^t-s_{x-1}^t=-1,+1$. 
    This is a direct consequence of the local conservation of particles: if the front of waiting particles increases, it means that during its propagation a particle have been taken out from a site, and vice versa.
    Also, we anticipate the CV $s_x^{t+1}\to s_{x-1}^{t+1}$ and have the independence 
    \begin{align*}
        P(s_{x-1}^{t+1}|s_{x-1}^{t}\cap s_{x}^{t}=s_{x-1}^{t}+1)&=P(s_{x-1}^{t+1}|s_{x-1}^{t})
    \end{align*}
    as the dynamic is directed and $s_{x-1}^{t+1}$ can only depend on $z(x')$ for $x'\leq x-1$.
    We can now recast Equation \eqref{eq: condition of decreasing} into
    \begin{align*}
       \sum_{s_{x-1}^{t},~s_{x-1}^{t+1}}&\big[(s_{x-1}^{t+1}-1)p+s_{x-1}^{t+1}q
        -(s_{x-1}^{t+1}p + (s_{x-1}^{t+1}+1)q)\big]P(s_{x-1}^{t+1}|s_{x-1}^{t})P(s_{x-1}^{t})\\
        &=-\sum_{s_{x-1}^{t},~s_{x-1}^{t+1}=1}^{x}P(s_{x-1}^{t+1}\cap s_{x-1}^{t})<0
    \end{align*}
    since all probabilities are positives.
    We notice that this term corresponds exactly to the probability for the two successive avalanches to make at least one toppling each at site $x-1$, starting from the stationary state. 
    Together with Equation \eqref{eq: recursion <s(x,t)s(x,t+1)>}, this concludes the proof for $\tau=1$.
    
    In the general case $\tau\geq 1$, the exact same reasoning applies, apart from the probabilities in Equations \eqref{eq: set conditional prob tau=1} for which we need to work a bit more to get an expression. 
    Following the steps as the one leading to Equation \eqref{eq: recursion <s(x,t)s(x,t+1)>}, one gets
    \begin{multline}\label{eq: recursion <s(x,t)s(x,t+tau)>}
        \avrg{s(x,t)s(x,t+\tau)} = \avrg{s(x-1,t)s(x-1,t+\tau)}\\
        +\alpha \sum_{s_{x-1}^{t},~s_x^{t+\tau}}s_x^{t+\tau} \big[P(s_x^{t+\tau}|s_{x-1}^{t}\cap s_{x}^{t}=s_{x-1}^{t}+1)\\
        -P(s_x^{t+\tau}|s_{x-1}^{t}\cap s_{x}^{t}=s_{x-1}^{t}-1)\big]P(s_{x-1}^{t})
    \end{multline}
    The problem is now that between the first and the $\tau^{th}$ avalanche, there is an ordered sequence $(s^{t'})_{t'}$ for $t'\in\intrange[1,\tau-1]$ of avalanches.
    In fact we can treat the intermediate $\tau-1$ avalanches as a unique super avalanche $\tilde{s}$, using the abelianess of the rules. 
    The only relevant information left is to know if this super avalanche has toppled at $x-1$ or not.
    This contribution already appeared in the work on the density $2$-points correlation function of the previous section. 
    We define a sequence of Bernoulli RV $\big(\epsilon(s_{x-1}^t,\tau-1)\big)_{s_{x-1}^t}$ which says if $\tilde{s}$ involved a toppling at $x-1$ (success event) or not (failure event) respectively \emph{knowing that $s_{x-1}^t$ topplings occurred along the first avalanche and $\tau-1$ intermediate avalanches are involved} in the super avalanche. 
    We denote by $\theta:=\theta(s_{x-1}^t,\tau-1)$ the parameter of success.
    For fixed $s_{x-1}^t$, conditioned on the success event $\epsilon=\epsilon(s_{x-1}^t,\tau-1)=1$, the local state after $\tilde{s}$ at $x$ is the stationary one.
    Conditioned on the failure event $\epsilon=0$, the local state after $\tilde{s}$ at $x$ is still given by the knowledge of the first avalanche, precisely on the value of $s_x^t-s_{x-1}^t$.
    This leads naturally to
    \begin{align*}
        P(s_{x}^{t+\tau}&|s_{x-1}^{t}\cap s_{x}^{t}=s_{x-1}^{t}+1\cap s_{x-1}^{t+\tau}=s_x^{t+\tau}-\delta^{t+\tau})\\
        &=\theta P(s_{x}^{t+\tau}|s_{x-1}^{t}\cap s_{x}^{t}=s_{x-1}^{t}+1 \cap s_{x-1}^{t+\tau}=s_x^{t+\tau}-\delta^{t+\tau}, \epsilon = 1)\\
        &+(1-\theta)P(s_{x}^{t+\tau}|s_{x-1}^{t}\cap s_{x}^{t}=s_{x-1}^{t}+1\cap s_{x-1}^{t+\tau}=s_x^{t+\tau}-\delta^{t+\tau}, \epsilon = 0)\\
        &=\left\{\begin{array}{cc}
            (1-\theta)p+\theta qp &  \text{if}~\delta^{t+\tau}= -1 \\
            (1-\theta)q+\theta (1-2pq) &  \text{if}~\delta^{t+\tau}= 0 \\
            \theta pq &  \text{if}~\delta^{t+\tau}= 1
        \end{array}\right.
    \end{align*}
    and also 
    \begin{align*}
        P(s_{x}^{t+\tau}&|s_{x-1}^{t}\cap s_{x}^{t}=s_{x-1}^{t}-1\cap s_{x-1}^{t+\tau}=s_x^{t+\tau}-\delta^{t+\tau})\\
        &=\left\{\begin{array}{cc}
            \theta qp &  \text{if}~\delta^{t+\tau}= -1 \\
            (1-\theta)p+\theta (1-2pq) &  \text{if}~\delta^{t+\tau}= 0 \\
            (1-\theta)q+\theta pq &  \text{if}~\delta^{t+\tau}= 1
        \end{array}\right.
    \end{align*}
    The two cases $\epsilon =0,1$ are separated formally in the above expressions taking respectively the term proportional to $(1-\theta)$ and $\theta$.
    Also, one should not forget the conditioning of $\epsilon$ and $\theta$ upon $\tau-1$ and $s_{x-1}^{t}$, a dependency that must be kept until the end of the computation.
    One can verify that for each cases, the sum of the probabilities over all the events gives the normalisation condition, i.e. $1$.
    Again, as was already true for $\tau=1$, we have the independence relation
    \begin{align*}
        P(s_{x-1}^{t+\tau}|s_{x-1}^{t}\cap s_{x}^{t}=s_{x-1}^{}+1\cap \epsilon)&=P(s_{x-1}^{t+\tau}|s_{x-1}^{t}\cap \epsilon)
    \end{align*}
    We can then express 
    \begin{align*}
        \sum_{s_{x-1}^{t},~s_x^{t+\tau}}&s_x^{t+\tau} \big[P(s_x^{t+\tau}|s_{x-1}^{t}\cap s_{x}^{t}=s_{x-1}^{t}+1)
        -P(s_x^{t+\tau}|s_{x-1}^{t}\cap s_{x}^{t}=s_{x-1}^{t}-1)\big]P(s_{x-1}^{t})\\
        &=\sum_{s_{x-1}^{t},~s_x^{t+\tau}}\theta(s_{x-1}^t,\tau-1)s_x^{t+\tau} \big[P(s_x^{t+\tau}|s_{x-1}^{t}\cap s_{x}^{t}=s_{x-1}^{t}+1\cap\epsilon = 1)
        \\
        &-P(s_x^{t+\tau}|s_{x-1}^{t}\cap s_{x}^{t}=s_{x-1}^{t}-1\cap\epsilon = 1)\big]P(s_{x-1}^{t})\\
        &+\sum_{s_{x-1}^{t},~s_x^{t+\tau}}(1-\theta(s_{x-1}^t,\tau-1))s_x^{t+\tau} \big[P(s_x^{t+\tau}|s_{x-1}^{t}\cap s_{x}^{t}=s_{x-1}^{t}+1\cap\epsilon = 0)
        \\
        &-P(s_x^{t+\tau}|s_{x-1}^{t}\cap s_{x}^{t}=s_{x-1}^{t}-1\cap\epsilon = 0)\big]P(s_{x-1}^{t})
    \end{align*}
    After the CV $s_{x}^{t+\tau}\to s_{x-1}^{t+\tau}$ the first term with $\epsilon = 1$ gives
    \begin{align*}
        \sum_{s_{x-1}^{t},~s_{x-1}^{t+\tau}}&\theta(s_{x-1}^t,\tau-1)\big[(s_{x-1}^{t+\tau}-1)pq+s_{x-1}^{t+\tau}(1-2pq)+(s_{x-1}^{t+\tau}+1)pq\\
        &-((s_{x-1}^{t+\tau}-1)pq+s_{x-1}^{t+\tau}(1-2pq)+(s_{x-1}^{t+\tau}+1)pq\big] P(s_{x-1}^{t+\tau}|s_{x-1}^{t}\cap \epsilon=1)P(s_{x-1}^{t})=0
    \end{align*}
    and the second one depending on $\epsilon=0$ gives
    \begin{align*}
        \sum_{s_{x-1}^{t},~s_{x-1}^{t+\tau}}&(1-\theta(s_{x-1}^t,\tau-1))\big[(s_{x-1}^{t+\tau}-1)p+s_{x-1}^{t+\tau}q-(s_{x-1}^{t+\tau}p+(s_{x-1}^{t+\tau}+1)q)\big]\times \\
        &P(s_{x-1}^{t+\tau}|s_{x-1}^{t}\cap \epsilon=0)P(s_{x-1}^{t})\\
        &=-\sum_{s_{x-1}^{t},~s_{x-1}^{t+\tau}\geq 1}(1-\theta(s_{x-1}^t,\tau-1))P(s_{x-1}^{t+\tau}|s_{x-1}^{t}\cap \epsilon=0)P(s_{x-1}^{t})<0
    \end{align*}
    Injecting the expression in Equation \eqref{eq: recursion <s(x,t)s(x,t+tau)>} finishes the proof.
\end{proof}
We confirm the exactness of the recurrence \eqref{eq: precise recursion <s(x,t)s(x,t+tau)>} using a direct numerical approach.
This is done by verifying that a brut evaluation, using the definition of $\avrg{s(x,t)s(x,t+\tau)}$, gives the same result as the recursive relation \eqref{eq: precise recursion <s(x,t)s(x,t+tau)>}, for which the first term $\avrg{s(1,t)s(1,t+\tau)}$ must be calculated beforehand. 
For example we get from the Proposition \ref{lemma: <s(x,t)s(x,t+1)> decreases with x} the relation $\avrg{s(2,t)s(2,t+1)}=\avrg{s(1,t)s(1,t+1)}-pq(p^2+pq^2+q^3)$ which is indeed verified for all $p\in[0,1]$.
Remark also that the negative term appearing in Equation \eqref{eq: precise recursion <s(x,t)s(x,t+tau)>} cannot be easily computed as the conditional probability $P(s_{x-1}^{t+\tau}|s_{x-1}^{t}\cap \epsilon=0)$ asks to perform the $\tau^{th}$ avalanche on a correlated state, which seems a hard task even using the RW picture.
This contrast with the result on the density correlations, in particular with Equation \eqref{eq: scaling correlations z(x,t)}.
In fact, we can interpret recursion \eqref{eq: precise recursion <s(x,t)s(x,t+tau)>} using the operator language already introduced in Proposition \ref{prop: finite time correlations}.

\begin{corollary}\label{corollary: matrix expression local avl anticor}
Denote by $W=W_x^0+W_x^{\geq 1}$ the transition matrix which we decompose into  
\begin{align}
    W_x^{0}(r\to r') &= \delta(1+\sum_x z^r(x),\sum_xz^{r'}(x))W(r\to r')\\
    W_x^{\geq 1}(r\to r') &= (1-\delta(1+\sum_x z^r(x),\sum_x z^{r'}(x)))W(r\to r')
\end{align}
$W_x^0$ correspond to all the transitions of $W$ where the triggered avalanche front stops before or at site $x$ and $W_x^{\geq 1}$ to all the other transitions of $W$ with at least one toppling at $x$.
In that case the recursion on the avalanche $2$-points correlation \eqref{eq: precise recursion <s(x,t)s(x,t+tau)>} can be written as \begin{equation}
    \avrg{s(x,t)s(x,t+\tau)} = \avrg{s(x-1,t)s(x-1,t+\tau)}\\
        -\alpha \bra[\Id] W_{x-1}^{\geq 1} (W_{x-1}^0)^{\tau-1} W_{x-1}^{\geq 1}\ket[\psi]
\end{equation}
\end{corollary}
\begin{proof}
    This result is a direct consequence of Equation \eqref{eq: precise recursion <s(x,t)s(x,t+tau)>} which can be recast into 
    \begin{align*}
        \avrg{s(x,t)s(x,t+\tau)} = &\avrg{s(x-1,t)s(x-1,t+\tau)}\\
        &-\alpha \sum_{s_{x-1}^{t},~s_{x-1}^{t+\tau}=1}^{x}P(s_{x-1}^{t}\cap\epsilon(\tau-1, s_{x-1}^{t})=0\cap s_{x-1}^{t+\tau})
    \end{align*}
    The rest follows from the definition of $P(s_{x-1}^{t}\cap\epsilon(\tau-1, s_{x-1}^{t})=0\cap s_{x-1}^{t+\tau})$ which asks, in causal order, to perform all avalanches at time $t$ that topple at least once at $x-1$, then perform $\tau-1$ avalanches which do not topple at $x-1$, and finally perform the avalanches at time $t+\tau$ that topple at least once at $x-1$.
\end{proof}
We can remark that the filtered transition matrix $W_{0,x}$ is a nilpotent matrix of index $x+1$ i.e. $(W_{0,x})^{x}> (W_{0,x})^{x+1}=0$.
This must indeed be true in light of Proposition \ref{prop: finite time correlations}.
From the previous results, we can now conclude on the sign of the avalanche $2$-points correlation function.

\begin{prop}\label{prop: local avl anticor}
The $2$-points correlations for the local number of topplings satisfies
\begin{equation}
    \avrg{s(x,t)s(x,t+\tau)}-\avrg{s(x,t)}\avrg{s(x,t+\tau)}\left\{\begin{array}{cc}
        < 0 & \text{if}~\tau\leq x \\
        = 0 & \text{if}~\tau\geq x+1
    \end{array} \right.
\end{equation}
\end{prop}
\begin{proof}
    From Proposition \ref{prop: finite time correlations} we have 
    \begin{equation*}
        \avrg{s(x,t)s(x,t+\tau)}=1
    \end{equation*}
    for $\tau \geq x+1$. 
    Lemma \ref{lemma: <s(x,t)s(x,t+1)> decreases with x} tell us that $\avrg{s(x,t)s(x,t+\tau)}$ decreases strictly with $\tau$ as long as the second term on the r.h.s. of Equation \eqref{eq: precise recursion <s(x,t)s(x,t+tau)>} is non zero. 
    
    We prove the strict decreasing by showing that the event $\epsilon(s_{x-1}^t,\tau-1)=0$ appearing with the probability $1-\theta(s_{x-1}^t,\tau-1)$ in Equation \eqref{eq: precise recursion <s(x,t)s(x,t+tau)>} is non zero for at least one $s_{x-1}^t$ as long as $1\leq \tau\leq x$.
    It suffices to consider the initial avalanche which empties the configurations of all of their particles, leaving the configuration $\zcfg{0^L}$ after its propagation. 
    By construction this can only occur with $s_{x-1}^{t}>0$ topplings at site $x-1$. 
    Selecting now the intermediate super avalanche, starting from $a_1^{\tau-1}\zcfg{0^L}$, which does the transition to $\zcfg{1^{x-1},0^{L-x+1}}$, where $\zcfg{1^{k}}:=\bigotimes_{i=1}^k\zcfg{1}$ from consecutive sites, leads to $1-\theta(s_{x-1}^t,\tau-1)>0$.
    In the end we get the chain of inequalities
    \begin{align*}
        1=\avrg{s(\tau-1,t)s(\tau-1,t+\tau)}>\avrg{s(\tau,t)s(\tau,t+\tau)}>\avrg{s(\tau+1,t)s(\tau+1,t+\tau)}>...
    \end{align*}
    which translates into 
     \begin{align*}
        0=\avrg{s(\tau-1,t)s(\tau-1,t+\tau)}-&\avrg{s(\tau-1,t)}\avrg{s(\tau-1,t+\tau)}\\
        &>\avrg{s(\tau,t)s(\tau,t+\tau)}-\avrg{s(\tau,t)}\avrg{s(\tau,t+\tau)}>...
    \end{align*}
    as we have $\avrg{s(x,t)}\avrg{s(x,t+\tau)}=1$ for all $x\geq 1$ and $\tau\geq 0$. 
\end{proof}

\begin{prop}
For any system size $L$, the avalanche time correlation satisfies 
\begin{equation}
    \avrg{s(t) s(t+\tau)}- \avrg{s(t)}\avrg{s(t+\tau)}\left\{\begin{array}{cc}
        < 0 & \text{if}~\tau\leq L \\
        = 0 & \text{if}~\tau\geq L+1
    \end{array} \right.
\end{equation}
For $\tau\leq L$ we have precisely
\begin{multline}\label{eq: exact <s(t)s(t+tau)>}
    \avrg{s(t)s(t+\tau)}= L^2\avrg{s(1,t)s(1,t+\tau)}\\
    -\alpha L\sum_{x=1}^{L-1}(L-x)\bigg[\sum_{s_{x}^{t},~s_{x}^{t+\tau}=1}^{x+1}P(s_{x}^{t}\cap\epsilon(\tau-1, s_{x}^{t})=0\cap s_{x}^{t+\tau})\bigg]
\end{multline}
\end{prop}
\begin{proof}
    It is very much the same argument as in Proposition \ref{prop: persistance global density} where the underlying linearity of the averaging operation solves the problem. 
    By construction we have 
    \begin{align*}
        \avrg{s(t) s(t+1)} &= \sum_{r,r',r''\in\recur}P(r)P(a_1r\to r')s(a_1r\to r')P(a_1r'\to r'')s(a_1r'\to r'') \\
        &=\sum_{r,r',r''\in\recur}P(r)P(a_1r\to r')P(a_1r'\to r'') \sum_{x,x'}s_x(a_1r\to r')s_{x'}(a_1r'\to r'')\\
        &=\sum_{x,x'}\sum_{r,r',r''\in\recur}P(r)P(a_1r\to r')P(a_1r'\to r'')s_x(a_1r\to r')s_{x'}(a_1r'\to r'')\\
        &=\sum_{x,x'}\avrg{s(x,t) s(x',t+1)}
    \end{align*}
    We are left with 
    \begin{align*}
        \avrg{s(t) s(t+\tau)}- \avrg{s(t)}\avrg{s(t+\tau)}&= \sum_{x,x'}(\avrg{s(x,t) s(x',t+\tau)}-\avrg{s(x,t)}\avrg{s(x',t+\tau)})\\
        &= L\sum_{x=1}^{L}(\avrg{s(x,t)^2}-\avrg{s(x,t)}\avrg{s(x,t+\tau)})<0
    \end{align*}
    in virtue of Propositions \ref{prop: min s(x,t) s(x',t+tau)} and the Lemma \ref{prop: local avl anticor}. 
    Finally, we have 
    \begin{align*}
        \avrg{s(x,t)s(x,t+\tau)} &= \avrg{s(1,t)s(1,t+\tau)} \\
        &-\alpha \sum_{x'=1}^{x-1}\bigg[\sum_{s_{x'}^{t},~s_{x'}^{t+\tau}=1}^{x'+1}P(s_{x'}^{t}\cap\epsilon(\tau-1, s_{x'}^{t})=0\cap s_{x'}^{t+\tau})\bigg]
    \end{align*}
    iterating the recursive relation \eqref{eq: recursion <s(x,t)s(x,t+tau)>}.
    Summing over $x$ leads to Equation \eqref{eq: exact <s(t)s(t+tau)>}.
\end{proof}
The recursion \eqref{eq: recursion <s(x,t)s(x,t+tau)>} provides direct control over the scaling in space of the correlations.
On the other hand, it is not clear to us how to get control on the scaling with increasing $\tau$.
We tried a few unsuccessful approaches.
The main obstacles we identified to perform this analysis come from comparing the magnitude of different contributions of opposite signs, for which we lack estimates, and/or evaluating the response function of correlated states, which again is not a simple task as the computation depends on the details of the states.
Motivated by our numerical investigation and the regularity observed we propose the following conjecture.
\begin{conjecture}\label{conjecture:<s(x,t)s(x,t+tau)> increases with tau}
    $\avrg{s(x,t)s(x,t+\tau)}$ is an increasing function in the regime $1\leq \tau \leq L$.
\end{conjecture}
\section{Conclusion}

In this paper, we characterized the $2$-points correlation function of two observable of the $1$D Directed Stochastic Sandpile model: the density and the number of toppling in the avalanches.
In particular, for both observable we found recursive relations \eqref{eq: scaling correlations z(x,t)}, \eqref{eq: precise recursion <s(x,t)s(x,t+tau)>} which have been powerful tools for their study.

We proved the persistence of the particle density which emerges as a trade-off.
The stationary state condition imposes the dissipation of exactly one particle, on average, per particle added to the system.
However, the natural spreading/extension of the avalanches does not scale as fast as the system size resulting in the accumulation of particles in the system.
This process leaves necessarily a subpart of the system unperturbed by the avalanche, precisely all sites at the right of the last site that has experienced a toppling in the avalanche. 
This absence of perturbation is exactly the contribution that develops the persistence of the local, and a fortiori global, density of particles.

For the avalanche $2$-points correlation, a careful analysis allowed to prove its negative sign up to a certain time $\tau^*$ at and after which it becomes null. 
We proved that this correlation depends only, at a fixed time, on the leftmost site of the two evaluated.
In fact, the time $\tau^*$ that separates the anticorrelated to the no correlation regime is exactly given by the position of the leftmost site plus one. 
The anticorrelation can be interpreted as the fact that avalanches create defaults in the system which are strictly recovered after a finite number of intermediate avalanches in between the initial and the last avalanches. 
Along this intermediate sequence of avalanches, the system gradually repairs and manage to recover the stationary state in finite time. 
To our knowledge, this is the first time that this kind of exact results on the space-time correlations of the DSS are derived.

The directedness has been of a great importance in the analysis of the model.
In particular, when looking at properties at a site $x$, it allowed to reduce the relevant information to depend only on the sites at the left of $x$.
Besides, it allowed to treat in a causal way the conditioning of the events.
Similarly, the conservation laws in the system and the complete irreversibly of the stabilisation process (no loop occurs along one avalanche) allowed to exploit, in a controlled way, the equivalence between one toppling and the knowledge of having one more grain on the nearest right site. 
Those are the main ingredients that allowed to define the simple recursion formula for both density and avalanche correlations.

Besides, we left a certain number of questions unanswered.
We did not manage to estimate the scaling of the avalanche $2$-points correlation function.
Nonetheless, a straightforward numerical implementation of the dynamics for small system sizes suggests a rather simple scaling form, which resembles, in this respect, the scaling form obtained for the density $2$-points correlation.
This problem could be solved if an explicit or asymptotic form for the negative term in Equation \eqref{eq: precise recursion <s(x,t)s(x,t+tau)>} was found, which seems an interesting open question. 
On another note, we believe that many more properties could be accessible in an exact fashion, exploiting some hidden symmetries of the model. 
This is well illustrated by the avalanche probability distribution in the stationary state which is exactly the same before and after switching the microscopic probabilities of the model $p\leftrightarrow q$ \cite{pruessner_exact_2004}. 
We have shown this property to be also true for the $2$-points correlation function of the particle density with \eqref{eq: scaling correlations z(x,t)}.
However, proving that this property holds for the avalanche $2$-points correlation function, as have been suggested by some preliminary numerical studies of our own, seems more involved and is out of the scope of this paper.
Supported again by the numerics, we also conjectured (Conjecture \ref{conjecture:<s(x,t)s(x,t+tau)> increases with tau}) that the avalanche $2$-points correlation is increasing with $\tau$, which at the heuristic level can be understood as the gradual repairing of the system from an initial perturbation.

\printbibliography[title={Bibliography}]
\end{document}